\documentclass[a4paper,11pt]{article}
\usepackage{pos}
\usepackage{mathrsfs}
\usepackage{dsfont}
\usepackage{axodraw2}
\makeatletter
\g@addto@macro\bfseries{\boldmath}
\makeatother	

\makeatletter
\newcommand*\rel@kern[1]{\kern#1\dimexpr\macc@kerna}
\newcommand*\widebar[1]{%
  \begingroup
  \def\mathaccent##1##2{%
    \rel@kern{0.8}%
    \overline{\rel@kern{-0.8}\macc@nucleus\rel@kern{0.2}}%
    \rel@kern{-0.2}%
  }%
  \macc@depth\@ne
  \let\math@bgroup\@empty \let\math@egroup\macc@set@skewchar
  \mathsurround\z@ \frozen@everymath{\mathgroup\macc@group\relax}%
  \macc@set@skewchar\relax
  \let\mathaccentV\macc@nested@a
  \macc@nested@a\relax111{#1}%
  \endgroup
}
\makeatother

\let\Re\relax% Set equal to \relax so that LaTeX thinks it's not defined
\let\Im\relax% Set equal to \relax so that LaTeX thinks it's not defined
\DeclareMathOperator{\Re}{Re}
\DeclareMathOperator{\Im}{Im}

\newcommand\Tstrut{\rule{0pt}{2.9ex}}         % "top" strut
\newcommand\Bstrut{\rule[-1.2ex]{0pt}{0pt}}   % "bottom" strut
\newcommand\TBstrut{\Tstrut\Bstrut}    
\newenvironment{Eqnarray}%
     {\arraycolsep 0.14em\begin{eqnarray}}{\end{eqnarray}}
\newcommand{\beq}{\begin{equation}}
\newcommand{\eeq}{\end{equation}}
\newcommand{\beqa}{\begin{Eqnarray}}
\newcommand{\eeqa}{\end{Eqnarray}}
\def\lsim{\mathrel{\raise.3ex\hbox{$<$\kern-.75em\lower1ex\hbox{$\sim$}}}}
\def\gsim{\mathrel{\raise.3ex\hbox{$>$\kern-.75em\lower1ex\hbox{$\sim$}}}}
\def\vev#1{\langle #1 \rangle}
\def\pht{\phantom{i}}
\def\phm{\phantom{-}}
\def\half{\tfrac12}
\def\nn{\nonumber}
\def\eq#1{eq.~(\ref{#1})}

\def\eqs#1#2{eqs.~(\ref{#1}) and~(\ref{#2})}

\def\eqst#1#2{eqs.~(\ref{#1})--(\ref{#2})}

\title{A natural mechanism for a SM-like Higgs boson in the 2HDM without decoupling}
\author*[a]{Howard E.~Haber}

\affiliation[a]{Santa Cruz Institute for Particle Physics,\\
University of California, Santa Cruz, CA 95064 USA}

\emailAdd{haber@scipp.ucsc.edu}

\abstract{The properties of the Higgs boson discovered at the Large Hadron Collider are very well described by the Standard Model (SM).  Thus, any
theory that invokes an extended Higgs sector must explain why the neutral scalar observed at the LHC so closely resembles the SM Higgs boson.  In this talk,
I review the Higgs alignment limit, in which one neutral scalar state of the Higgs sector is SM-like.  An approximate Higgs alignment can be achieved 
``naturally'' either via decoupling or via an approximate symmetry.  Using the two-Higgs doublet model as a prototype for an extended Higgs sector,
I examine the symmetries of the scalar potential and their soft breakings that may be responsible for the SM-like properties of the observed Higgs boson,
and I demonstrate how to extend such (softly-broken) symmetries to the Yukawa sector of the model.}

\FullConference{%
  7th Symposium on Prospects in the Physics of Discrete Symmetries (DISCRETE 2020-2021)\\
  29th November - 3rd December 2021\\
 Bergen, Norway}

\begin{document}
\maketitle

\section{Introduction}

Nearly ten years after the initial discovery of the Higgs boson, 
the LHC Higgs data have already achieved a precision that implies that the properties of the observed neutral scalar closely approximate those of the Standard Model (SM) Higgs boson 
to within an accuracy that is typically in the range of $10\%$--$20\%$ depending on the observable~\cite{ATLAS:2022vkf,CMS:2022dwd}. 
One possible conclusion is that the scalar sector responsible for electroweak symmetry breaking is of minimal form, resulting in one physical neutral spin-zero state that can be identified with the scalar observed at the LHC.

Nevertheless, given the current precision of the Higgs data, the possibility that the Higgs sector contains more that one physical scalar cannot be excluded.
It is noteworthy that the structure of the Standard Model is far from being of minimal form. For example, there are three generations of quarks and leptons whereas one generation would have been sufficient. The SM gauge group is SU(3)$\times$SU(2)$\times$U(1), which is again of a non-minimal form.  So why shouldn't the scalar sector be non-minimal as well?
If a non-minimal scalar sector exists, one obvious question to ask is: why is the observed Higgs boson SM-like?

Consider a non-minimal scalar sector in which all scalars (apart from the SM-like Higgs boson) are very heavy, say, with masses above some heavy scale $M\gsim 1$~TeV.   
One can then formally integrate out all the heavy scalar states from the theory.  At scales below $M$, the scalar sector of the resulting low energy effective theory consists of 
one complex Higgs doublet, which contains the three Goldstone fields that provide the masses for the $W^\pm$ and $Z$ gauge bosons and one physical
neutral scalar that coincides precisely with the SM Higgs boson.  This is known as the decoupling limit of the extended Higgs sector, and provides
a natural explanation for the SM-like nature of the observed Higgs boson~\cite{Haber:1989xc,Gunion:2002zf,Haber:2006ue}.\footnote{``Natural'' is a loaded term.  In this talk, I will not provide an explanation for
why the scale of electroweak symmetry breaking, $v\simeq 246$~GeV, is so much smaller than, say, the Planck scale.  This requires one unnatural fine-tuning of the Higgs sector
parameters, whose origin will not be addressed in this work.  (For a review of naturalness, see Ref.~\cite{Craig:2022uua}.)}

One disadvantage of the decoupling limit is that the additional scalar states beyond the observed Higgs boson may be difficult (or impossible) to discover at the LHC due
to insufficient energy of the collider and/or to SM backgrounds that overwhelm any potential signal.  Thus, in this talk, I shall focus on the possible existence of a SM-like
Higgs boson, where some of the additional scalar states are not significantly heavier than the Higgs boson of mass 125 GeV and thus are potentially accessible at the LHC in future Higgs studies.

A typical feature of an extended Higgs sector is the presence of a neutral scalar field, $\varphi$, whose tree-level interactions with the gauge bosons, fermions and its self-interactions are precisely those of the SM Higgs field.  Generically, $\varphi$ can mix with other neutral scalar fields of the extended Higgs sector.  The physical scalars of the model are then obtained by diagonalizing the neutral scalar squared mass matrix.  If it turns out that $\varphi$ is an approximate eigenstate of the squared-mass matrix (due to suppressed mixing with other neutral scalar fields of the extended Higgs sector), then one of the physical neutral scalar states will be SM-like~\cite{Ginzburg:2001wj,Gunion:2002zf}.  
The limit of zero mixing is called the \text{Higgs alignment limit}~\cite{Craig:2013hca,Asner:2013psa,Carena:2013ooa,Haber:2013mia,Carena:2015moc}.  In this case, the other physical scalars of the model may or may not be significantly heavier than the SM Higgs boson. In particular, the decoupling limit [in which the mixing is suppressed by $\mathcal{O}(v^2/M^2)$] is a special case of the Higgs alignment limit.

As an example, consider an extended Higgs sector with $n$ hypercharge-one Higgs doublets $\Phi_{i}$ and $m$ additional neutral singlet Higgs fields $\phi^0_{j}$.
After minimizing the scalar potential, we assume that only the neutral Higgs fields acquire vacuum expectation values (in order to preserve U(1)$_{\mathrm{EM}})$,
\beq
\left\langle\Phi_{i}^{0}\right\rangle=\frac{v_{i}}{\sqrt{2}}\,, \qquad \bigl\langle{\phi_{j}^{0}}\bigr\rangle=x_{j}\,,
\eeq
where $v^{2} \equiv \sum_{i}\left|v_{i}\right|^{2}=4 m_{W}^{2} / g^{2}=(246~\mathrm{GeV})^{2}$.

Define new linear combinations of the hypercharge-one doublet Higgs fields (the so-called Higgs basis~\cite{Georgi:1978ri,Lavoura:1994yu,Lavoura:1994fv,Botella:1994cs,Branco:1999fs}). In particular,
\beq
H_{1}=\left(\begin{array}{c}
H_{1}^{+} \\[3pt]
H_{1}^{0}
\end{array}\right)=\frac{1}{v} \sum_{i} v_{i}^{*} \Phi_{i}\,, \qquad \text{where $\left\langle H_{1}^{0}\right\rangle=\frac{v}{\sqrt{2}}$}\,,
\eeq
and $H_{2}, H_{3}, \ldots, H_{n}$ are the other linear combinations of doublet scalar fields such that $\left\langle H_{i}^{0}\right\rangle=0$ for $i=2,3, \ldots, n)$.
That is $H_{1}^{0}$ is \textit{aligned} in field space with the direction of the Higgs vacuum expectation value (vev). Thus, if $\varphi\equiv\sqrt{2} \operatorname{Re} H_{1}^{0}-v$ is a mass-eigenstate, then the tree-level couplings of $\varphi$ to itself, to gauge bosons and to fermions are precisely those of the SM Higgs boson. This is the exact Higgs alignment limit.

To achieve the exact Higgs alignment limit, one must explain why $\varphi$ does not mix with other neutral scalar fields $H_i^0$ ($i=2,3,\ldots n$) and $\phi_j^0$.
As noted above, the mixing is naturally suppressed in the decoupling limit.  But, in the alignment limit without decoupling, the absence of mixing appears to require a fine-tuning of the scalar potential parameters.   The central question of this talk is whether there is a natural mechanism that can produce approximate Higgs alignment without decoupling.

\section{The Higgs alignment limit of the 2HDM}

Let us focus on the two-Higgs doublet model (2HDM) as a prototype for an extended Higgs sector.
Consider the 2HDM scalar potential in the $\Phi_1$--$\Phi_2$ basis (e.g., see Ref.~\cite{Branco:2011iw}),
\beqa
\mathcal{V}&\!=\!&m_{11}^2\Phi_1^\dagger\Phi_1+m_{22}^2\Phi_2^\dagger\Phi_2
-[m_{12}^2\Phi_1^\dagger\Phi_2+{\rm h.c.}]
+\half\lambda_1(\Phi_1^\dagger\Phi_1)^2+\half\lambda_2(\Phi_2^\dagger\Phi_2)^2
+\lambda_3(\Phi_1^\dagger\Phi_1)(\Phi_2^\dagger\Phi_2)
 \nn\\[5pt]
&&\quad
+\lambda_4(\Phi_1^\dagger\Phi_2)(\Phi_2^\dagger\Phi_1)+\left\{\half\lambda_5(\Phi_1^\dagger\Phi_2)^2
+\big[\lambda_6(\Phi_1^\dagger\Phi_1)
+\lambda_7(\Phi_2^\dagger\Phi_2)\big]
\Phi_1^\dagger\Phi_2+{\rm h.c.}\right\}\,. \label{pot}
\eeqa 
The scalar fields $\Phi_i$ are hypercharge one, weak isospin doublets. 
After minimizing the scalar potential, $\vev{\Phi_i^0}=v_i/\sqrt{2}$ (for $i=1,2$) with
$v\equiv (|v_1|^2+|v_2|^2)^{1/2}=246$~GeV.
One is free to rephase the fields $\Phi_1$ and $\Phi_2$ such that $v_1=vc_\beta$ and $v_2\equiv  vs_\beta e^{i\xi}$,
where $c_\beta\equiv \cos\beta$ and $s_\beta\equiv \sin\beta$, with $0\leq\beta\leq\pi/2$ and $0\leq\xi<2\pi$.
  
In the Higgs basis, new scalar doublet fields are defined,
\beq \label{hbasis}
\mathcal{H}_1=\begin{pmatrix}\mathcal{H}_1^+\\[3pt] \mathcal{H}_1^0\end{pmatrix}\equiv c_\beta \Phi_1+s_\beta e^{-i\xi}\Phi_2\,,
\qquad \mathcal{H}_2=\begin{pmatrix} \mathcal{H}_2^+\\[3pt] \mathcal{H}_2^0\end{pmatrix}\equiv e^{i\eta}\bigl(-s_\beta e^{i\xi}\Phi_1+c_\beta\Phi_2\bigr),
\eeq
such that 
 $\vev{\mathcal{H}_1^0}=v/\sqrt{2}$ and $\vev{\mathcal{H}_2^0}=0$.  The Higgs basis is uniquely defined up to an overall rephasing that is parameterized by the phase angle $\eta$~\cite{Boto:2020wyf}).
Rewriting $\Phi_1$ and $\Phi_2$ in terms of the Higgs basis fields and inserting the result into \eq{pot} yields the scalar potential in the Higgs basis,
\beqa
 \mathcal{V}&=& Y_1 \mathcal{H}_1^\dagger \mathcal{H}_1+ Y_2 \mathcal{H}_2^\dagger \mathcal{H}_2 +[Y_3 e^{-i\eta}
\mathcal{H}_1^\dagger \mathcal{H}_2+{\rm h.c.}]
\nn\\[4pt]
&&
+\half Z_1(\mathcal{H}_1^\dagger \mathcal{H}_1)^2+\half Z_2(\mathcal{H}_2^\dagger \mathcal{H}_2)^2
+Z_3(\mathcal{H}_1^\dagger \mathcal{H}_1)(\mathcal{H}_2^\dagger \mathcal{H}_2)
+Z_4( \mathcal{H}_1^\dagger \mathcal{H}_2)(\mathcal{H}_2^\dagger \mathcal{H}_1) \nn \\[4pt]
&&
+\left\{\half Z_5 e^{-2i\eta}(\mathcal{H}_1^\dagger \mathcal{H}_2)^2 +\big[{Z_6 e^{-i\eta} (\mathcal{H}_1^\dagger
\mathcal{H}_1)} +Z_7 e^{-i\eta} (\mathcal{H}_2^\dagger \mathcal{H}_2)\big] {\mathcal{H}_1^\dagger \mathcal{H}_2+{\rm
h.c.}}\right\}.
\eeqa
The parameters $Y_{1,2,3}$ and $Z_{1,2,\ldots,7}$ can be expressed in terms of the parameters of the scalar potential in the $\Phi_1$-$\Phi_2$ basis given in \eq{pot}.  
For example~\cite{Davidson:2005cw,Boto:2020wyf},
\beqa
Y_3 e^{i\xi}&=&\half(m_{22}^2-m_{11}^2)s_{2\beta}-\Re(m_{12}^2 e^{i\xi})c_{2\beta}-i\Im(m_{12}^2 e^{i\xi})\,,\label{whythree} \\[3pt]
Z_6 e^{i\xi}&=& -\half s_{2\beta}\left\{\lambda_1c^2_{\beta}-\lambda_2s_{\beta}^2-
\bigl[\lambda_3+\lambda_4+\Re(\lambda_5 e^{2i\xi})\bigr] c_{2\beta}-i\Im(\lambda_5 e^{2i\xi})\right\} \nn \\
&& \qquad  +c_\beta c_{3\beta}\Re(\lambda_6 e^{i\xi}) +s_\beta s_{3\beta}\Re(\lambda_7e^{i\xi}) 
 +ic_\beta^2\Im(\lambda_6 e^{i\xi})+is_{\beta}^2\Im(\lambda_7e^{i\xi})\,.\label{zeesix}
\eeqa
The scalar potential in the Higgs basis is minimized when the following two conditions are satisfied,
\beq \label{potmin}
Y_1=-\half Z_1 v^2\,,\qquad\quad Y_3=-\half Z_6 v^2\,.
\eeq

Exact Higgs alignment corresponds to the absence of $\mathcal{H}_1^0$--$\mathcal{H}_2^0$ mixing, which is achieved when $Z_6=0$ (and $Y_3=0$ via the scalar
potential minimum conditions).
I therefore pose the following question: how can one achieve the condition $Z_6=0$ naturally?  In practice, to be consistent with a SM-like Higgs boson
without decoupling, it is sufficient to demand that $|Z_6|\ll 1$.

The simplest way to guarantee that $Z_6=0$ is to introduce a $\mathbb{Z}_2$ symmetry in the Higgs basis such that $\mathcal{H}_2\to -\mathcal{H}_2$ is the only $\mathbb{Z}_2$-odd field of the two Higgs doublet extended Standard Model. In this case, $Z_6=0$ and the tree-level properties of $h\equiv \sqrt{2}\,\Re \mathcal{H}_1^0-v$ coincide with those of the SM Higgs boson.  
The $\mathbb{Z}_2$ symmetry is unbroken by the vacuum and thus remains exact.  Note that the fermions (which are  $\mathbb{Z}_2$-even) can only couple to $\mathcal{H}_1$, so that 
this model possesses Type-I Higgs-fermion interactions~\cite{Hall:1981bc}.   The model just described is known as the inert doublet mode (IDM), since the physical scalars that reside in $\mathcal{H}_2$ (consisting of a charged Higgs boson $H^+$ and two neutral scalars $H$ and $A$) cannot interact singly with the particles of the SM~\cite{Barbieri:2006dq,LopezHonorez:2006gr}.  
The lightest of the $\mathbb{Z}_2$-odd scalars is stable and thus is a candidate for dark matter~\cite{Cirelli:2005uq,Ma:2006km,LopezHonorez:2006gr,Hambye:2009pw,Arina:2009um,Goudelis:2013uca}.

In the IDM, the Higgs alignment limit is exact, and the tree-level couplings of $h$ are precisely those of the SM Higgs boson.  Deviations from SM behavior can only arise at loop level due to the effects of pairs of inert scalars that can appear in the loop.  For example, there would be a small deviation from the SM in the prediction of $\Gamma(h\to\gamma\gamma)$ mediated by a charged Higgs loop.

Suppose that experimental deviations from the SM Higgs boson properties are observed that can be attributed to a deviation from the exact Higgs alignment limit (i.e., $Z_6\neq 0$). 
The possibility of realizing such a scenario in future LHC running is considered in Refs.~\cite{Bernon:2015qea,Bernon:2015wef,Hou:2017hiw,Grzadkowski:2018ohf,Aiko:2020ksl,Kanemura:2021dez}.
One can accommodate such deviations in the IDM by perturbing the
model such that $Y_3$, $Z_6\neq 0$ [cf.~\eq{potmin}].    However, this would constitute a \textit{hard} breaking of the $\mathbb{Z}_2$ symmetry that governs the IDM. That is, in the framework of a perturbed IDM, there would be no natural explanation as to why $|Z_6|\ll 1$.   In the next section, I shall survey other global symmetries of the 2HDM scalar potential that yield exact Higgs alignment but allow for the possibility of soft symmetry breaking, thereby providing a natural mechanism for the presence of a SM-like Higgs boson~\cite{Haber:2021zva}.
Alternative approaches to symmetry based explanations of natural Higgs alignment can be found in Refs.~\cite{Dev:2014yca,Dev:2017org,Pilaftsis:2016erj,Benakli:2018vqz,Benakli:2018vjk,Lane:2018ycs,Eichten:2021qbm}.

\section{Global symmetries of the 2HDM bosonic sector}
\label{classify}

Two classes of global symmetries are considered below that can be imposed on the 2HDM scalar potential [\eq{pot}] and the gauge-covariant scalar kinetic energy terms:
Higgs family symmetries and generalized CP (GCP) symmetries.  Among the possible Higgs family symmetries are: 

$\mathbb{Z}_2:\hspace{27.3ex}
\Phi_1 \rightarrow \Phi_1,
\hspace{7.9ex}
\Phi_2 \rightarrow - \Phi_2$

$\Pi_2:
\hspace{27ex}
\Phi_1 \rightarrow \Phi_2,
\hspace{8ex}
\Phi_2 \rightarrow \Phi_1$

U(1)$_{\rm PQ}$\,\,\,({\rm Peccei-Quinn}~\cite{Peccei:1977hh}):
\hspace{4ex}
$\Phi_1 \rightarrow e^{-i \theta} \Phi_1,
\hspace{4ex}
\Phi_2 \rightarrow e^{i \theta} \Phi_2$

${\rm SO}(3)$:\hspace{25.1ex} \!\!$\Phi_a\to U_{ab}\Phi_b$\,,\,\qquad \!\!$U\in {\rm U}(2)/{\rm U}(1)_Y$

\noindent
where there is an implicit sum over the index $b$.
Note that the largest possible global symmetry of the bosonic sector of the 2HDM is U(2), which contains within it the gauged
hypercharge U(1)$_Y$ symmetry.  Removing the latter leaves a global symmetry that is isomorphic to SO(3).

Among the possible GCP symmetries are~\cite{Ivanov:2007de,Ferreira:2009wh,Ferreira:2010yh,Battye:2011jj}:

${\rm GCP1}:\hspace{6ex}
\Phi_1 \rightarrow \Phi_1^*,
\hspace{14ex}
\Phi_2 \rightarrow \Phi_2^*$

${\rm GCP2}:\hspace{6ex}
\Phi_1 \rightarrow \Phi_2^*,
\hspace{14ex}
\Phi_2 \rightarrow -\Phi_1^*$

${\rm GCP3}:\hspace{6ex}
\Phi_1 \rightarrow \Phi_1^*c_\theta+\Phi_2^*s_\theta,
\hspace{5ex}\!
\Phi_2 \rightarrow -\Phi_1^*s_\theta+\Phi_2^*c_\theta,\qquad \text{for $0<\theta<\half\pi$}$

\noindent
where $c_\theta\equiv \cos\theta$ and $s_\theta\equiv \sin\theta$.   Imposing the symmetries above constrains the parameters of \eq{pot} as shown in Table~\ref{symm}.
\begin{table}[h!]
\centering
\caption{Higgs family and GCP symmetries of the 2HDM scalar potential and the gauge-covariant scalar kinetic energy terms.  Constraints on the scalar potential parameters in the $\Phi_1$--$\Phi_2$ basis are shown~\cite{Ivanov:2007de,Ferreira:2009wh,Ferreira:2010yh,Battye:2011jj}. \label{symm}}
\begin{tabular}{|c|cccccccc|}
\hline 
\pht symmetry &  $m_{22}^2$ &\quad $m_{12}^2$ \quad & 
 $\lambda_2$ &  $\lambda_4$ &
$\Re\lambda_5$ &  $\Im\lambda_5$  & $\lambda_6$\quad  & \quad $\phm\lambda_7$ \quad  \TBstrut \\
\hline
$\mathbb{Z}_2$  &   & $0$
   &    &  & & 
   & $0$ & $\phm 0$ \\
$\Pi_2$    &$ m_{11}^2$ &\pht  real \pht &
    $ \lambda_1$ & &  
 & $0$ &  & $\phm\lambda_6^\ast$
\\
$\mathbb{Z}_2\otimes\Pi_2$  & $m_{11}^2$ & $0$ & $\lambda_1$ &&  & $0$ &  $0$ & $\phm 0$
\\
U(1)$_{\rm PQ}$  &  & $0$ 
 &  & &  
$0$ & $0$ & $0$ &  $\phm 0$ \\
U(1)$_{\rm PQ}\otimes\Pi_2$   & $m_{11}^2$ & $0$ & $\lambda_1$ &&  $0$ & $0$ & $0$ & $\phm 0$
\\
SO(3)  & $ m_{11}^2$ & $0$
   & $\lambda_1$   & \pht $\lambda_1 - \lambda_3$\pht  &
$0$ & $0$ & $0$ & $\phm 0$ \\
GCP1    & & real
 & &  &  
& $0$ &\pht  real \pht & \pht real \pht \\
GCP2   & $m_{11}^2$ & $0$
  & $\lambda_1$  &  &
&   &  & $- \lambda_6\pht$ \\
GCP3   & $m_{11}^2$ & $0$
   & $\lambda_1$  &  &
\pht  $\lambda_1 - \lambda_3 - \lambda_4$\pht  & $0$ & $0$ & $\phm 0$ \Bstrut \\
\hline
\end{tabular}
\end{table}

Not all the symmetries shown in shown in Table~\ref{symm} are inequivalent.  In particular, the $\Pi_2$ symmetry in the $\Phi_1$--$\Phi_2$ basis is equivalent to the $\mathbb{Z}_2$ symmetry in another scalar field basis.  Likewise $\mathbb{Z}_2\otimes\Pi_2$ is equivalent to GCP2 in another scalar field basis, and U(1)$_{\rm PQ}\otimes\Pi_2$ is equivalent to GCP3 in another scalar field basis~\cite{Ferreira:2009wh,Ferreira:2010yh,Haber:2021zva}.  That is, there are precisely six inequivalent global symmetries among the symmetries listed in Table~\ref{symm}.  One can prove that any global symmetry of the 2HDM bosonic sector (under the assumption of a scalar potential consisting of terms of dimension four or less) is equivalent to one of the six inequivalent global symmetries mentioned above~\cite{Ivanov:2007de,Ferreira:2009wh}.

Finally, we note the following exceptional region of the 2HDM parameter space first introduced in Ref.~\cite{Davidson:2005cw}
and subsequently designated by the acronym ERPS in Ref.~\cite{Ferreira:2009wh},
in which  $m_{11}^2=m_{22}^2$, $\lambda_1=\lambda_2$ and $\lambda_7=-\lambda_6$.  
The ERPS exhibits one of the following symmetries: 
 $\mathbb{Z}_2\otimes\Pi_2$, U(1)$_{\rm PQ}\otimes\Pi_2$ (or equivalently, GCP2, GCP3), or SO(3), manifestly realized in some scalar 
 field basis.

The ERPS has a number of remarkable properties. One of these results is exhibited by the theorem below (for which a simple proof is given in Ref.~\cite{Haber:2021zva}):

\textbf{Theorem}:  If  $\lambda_1=\lambda_2$ and $\lambda_7=-\lambda_6$, then these conditions hold in \textit{any} scalar field basis.  Moreover,
a basis of scalar fields (which is not unique) exists such that
$\lambda_6=\lambda_7=0$ and $\lambda_5\in\mathbb{R}$.

\section{Symmetry origin for (approximate) Higgs alignment in the 2HDM}

Consider the 2HDM scalar potential in the $\Phi_1$--$\Phi_2$ basis [\eq{pot}].  If $m_{11}^2=m_{22}^2$ and $m_{12}^2=0$, then \eq{whythree} yields $Y_3=0$.  By virtue of the scalar potential minimum conditions [\eq{potmin}], it follows that $Z_6=0$, which corresponds to the exact Higgs alignment limit.  A perusal of Table~\ref{symm} then shows that exact Higgs alignment arises if any one of the symmetries of the ERPS is satisfied.
Moreover, by virtue of the theorem quoted above, $Z_6=0$ in the ERPS implies that $Z_7=0$.  That is, the inert limit of $Y_3=Z_6=Z_7=0$ is satisfied.
However,  it is remarkable that in many cases, exact Higgs alignment is preserved even if the ERPS symmetries are softly broken.
In all such cases, exact Higgs alignment is achieved in the inert limit where $Y_3=Z_6=Z_7=0$.

The complete classification of 2HDM scalar potentials with exact Higgs alignment due to a symmetry was obtained in Ref.~\cite{Haber:2021zva} and
includes the IDM as well as scalar potentials that exhibit one of the ERPS symmetries.  However, additional
models of exact Higgs alignment can be constructed based on ERPS symmetries that are softly broken, where a residual symmetry remains unbroken by the vacuum. 
As an example, consider a $\mathbb{Z}_2\otimes\Pi_2$-symmetric scalar potential that is softly broken by setting $\Re m_{12}^2\neq 0$ (whereas, $m_{11}^2=m_{22}^2$ and  $\Im m_{12}^2=0$).  In this case, the minimization of the scalar potential 
yields $\cos 2\beta=\sin\xi=0$.  Note that in the $\Phi_1$--$\Phi_2$ basis, the $\mathbb{Z}_2$ symmetry is broken but the $\Pi_2$ symmetry remains intact.  Indeed, the residual $\Pi_2$ symmetry in the $\Phi_1$--$\Phi_2$ basis is equivalent to a $\mathbb{Z}_2$ symmetry in the Higgs basis.  Consequently, it follows that $Y_3=Z_6=Z_7=0$ 
[which can be checked in light of  \eqst{whythree}{potmin}] and exact Higgs alignment is preserved.

In the terminology employed in this talk, natural Higgs alignment corresponds to the existence of a scalar whose tree-level properties coincide with the SM Higgs boson as a consequence of a symmetry  (which is unbroken by the vacuum) rather than an artificial fine-tuning of the model parameters.   If the symmetry responsible for Higgs alignment is 
subsequently broken by soft symmetry-breaking terms, then the deviation from exact Higgs alignment can be \textit{naturally} small.   This is in keeping with the definition of naturally small parameters in the sense of `t Hooft, who argued that a small parameter is naturally small if the symmetry of the theory is enlarged in the limit where the 
soft symmetry-breaking parameter is set to zero~\cite{tHooft:1979rat}.

In Refs.~\cite{Dev:2014yca,Dev:2017org}, ``natural alignment'' is defined by requiring that $Y_3=Z_6=0$ is satisfied independently of the scalar potential minimum conditions.  In particular, $Y_3=Z_6=0$ 
given in \eqs{whythree}{zeesix}
must be satisfied independently of the values of $\beta$ and~$\xi$. In light of this stricter definition of ``natural,'' it follows that ``natural alignment'' implies 
that $m_{11}^2=m_{22}^2$, $m_{12}^2=0$, $\lambda_1=\lambda_2=\lambda_3+\lambda_4$, and $\lambda_5=\lambda_6=\lambda_7=0$, which are the conditions for the SO(3) 
symmetry\footnote{If one assumes real scalar potential parameters and $\xi=0$, then $Y_3=Z_6=0$ independently of $\beta$ yields
$m_{11}^2=m_{22}^2$, $m_{12}^2=0$, $\lambda_1=\lambda_2=\lambda_3+\lambda_4+\lambda_5$, and $\lambda_6=\lambda_7=0$, which are the conditions for the GCP3 symmetry exhibited in Table~\ref{symm}.  However, this is a basis-dependent result, since the same criteria applied to a U(1)$_{\rm PQ}\otimes\Pi_2$-symmetric scalar potential (which is equivalent to GCP3 in another scalar field basis) would not yield
``natural alignment.''}
exhibited in Table~\ref{symm} (and corresponds to the so-called maximally symmetric 2HDM of Ref.~\cite{Dev:2014yca}).

In this talk, I will \textit{not} employ the stricter version of natural Higgs alignment advocated in Refs.~\cite{Dev:2014yca,Dev:2017org}.  Since I am interested in scenarios where the deviation from exact Higgs alignment is naturally small in the sense of 't Hooft~\cite{tHooft:1979rat}, it is a useful exercise to classify the softly-broken symmetries of the ERPS in which $Z_6\neq 0$.   
A complete list of possible softly broken symmetries with $Z_6\neq 0$ (after imposing the scalar potential minimum conditions) can be found in Tables~\ref{tabalign1}, \ref{tabalign2} and \ref{tabalign3}.  The cases shown below where $m_A^2=0$ arise when the vacuum breaks a residual U(1)$_{\rm PQ}$ symmetry (thereby generating a massless Goldstone boson).  Such cases are phenomenologically untenable and can be excluded from further consideration.

\begin{table}[h!]
\caption{Scalar potentials in the $\Phi_1$--$\Phi_2$ basis with a softly-broken $\mathbb{Z}_2\otimes\Pi_2$ symmetry, where $\lambda\equiv\lambda_1=\lambda_2$, $R\equiv (\lambda_3+\lambda_4+\lambda_5)/\lambda$, $\Im\lambda_5=\lambda_6=\lambda_7=0$ and
$Z_6=-\half s_{2\beta} e^{-i\xi}\bigl\{[\lambda(1-R)+2\lambda_5\sin^2\xi]c_{2\beta}-i\lambda_5\sin 2\xi\bigr\}\neq 0$ [after making use of \eq{zeesix}].
\label{tabalign1}}
\begin{tabular}{|c|c|c|c|c|c|}
\hline
$\beta$ &  $\pht\sin 2\xi\pht$ & $m_{11}^2$, $m_{22}^2$ & $m_{12}^2$ & CP-violation? & comment \Bstrut
\\
\hline
$s_{2\beta}\neq 0$ & $\neq 0$ &  $m_{11}^2\neq m_{22}^2$ & complex & explicit &  $\Im\bigl[m_{12}^2\bigr]^2\neq 0$ \Tstrut \\
$s_{2\beta}\neq 0$ &  $\neq 0$ &  $m_{11}^2\neq m_{22}^2$  & $\pht\Im\bigl[m_{12}^2\bigr]^2=0\pht$ & spontaneous &  $0<|m_{12}^2|<\half\lambda_5 v^2 s_{2\beta}$   \\
$s_{2\beta}\neq 0$ & $\neq 0$ &  $m_{11}^2\neq m_{22}^2$  & $\Im\bigl[m_{12}^2\bigr]^2=0$& no &  $|m_{12}^2|>\half\lambda_5 v^2 s_{2\beta}$   \\
$c_{2\beta}= 0$ &  $\neq 0$ &  $\pht m_{11}^2= m_{22}^2\pht$  & complex & no &  $m_{12}^2\neq 0$ \\
$s_{2\beta}c_{2\beta}\neq 0$ & $0$ &  $m_{11}^2\neq m_{22}^2$  & $\Im\bigl[m_{12}^2\bigr]^2=0$ & no& \Bstrut  \\
\hline
\end{tabular}
\end{table}
\begin{table}[h!]
\caption{Scalar potentials  in the $\Phi_1$--$\Phi_2$ basis with a softly-broken U(1)$_{\rm PQ}\otimes\Pi_2$ symmetry, where $\lambda\equiv\lambda_1=\lambda_2$, $\lambda_5=\lambda_6=\lambda_7=0$, $R\equiv (\lambda_3+\lambda_4)/\lambda$, $\Im(m_{12}^2 e^{i\xi})=0$, 
and $Z_6=-\half\lambda s_{2\beta} c_{2\beta} e^{-i\xi}(1-R)\neq 0$ [after making use of \eq{zeesix}].  Note that  $m_A^2=2\Re(m_{12}^2 e^{i\xi})/s_{2\beta}\geq 0$.
The scalar potential and vacuum are CP-conserving.\label{tabalign2}}  
\begin{tabular}{|c|c|c|c|c|}
\hline
$\beta$ &  $m_{11}^2$, $m_{22}^2$ & $\pht \Re(m_{12}^2 e^{i\xi})\pht$ & $R$ & comment \Bstrut \\
\hline
$\pht s_{2\beta}c_{2\beta}\neq 0 \pht$ &  $\pht m_{11}^2\neq m_{22}^2\pht $ & $>0$ & $R\neq 1$ & \Tstrut \\
$\pht s_{2\beta}c_{2\beta}\neq 0 \pht$ &  $\pht m_{11}^2= m_{22}^2\pht $ & $> 0$ & $R> 1$ & $m_A^2=\half\lambda v^2(R-1)$ \\
$\pht s_{2\beta}c_{2\beta}\neq 0 \pht$ &  $\pht m_{11}^2\neq m_{22}^2\pht $ & $0$ & $|R|<1$ &   $m_A^2=0$  \Bstrut \\
\hline
\end{tabular}
\end{table}
 \begin{table}[h!]
 \caption{Scalar potentials in the $\Phi^\prime_1$--$\Phi^\prime_2$ basis with a softly-broken GCP3 symmetry, with $\lambda^\prime\equiv\lambda_1^\prime=\lambda_2^\prime$,
 $\Re\lambda_5^\prime=\lambda_1^\prime-\lambda_3^\prime-\lambda_4^\prime$, $\Im\lambda_5^\prime=\lambda_6^\prime=\lambda_7^\prime=0$, and $Z_6=i\lambda_5^\prime s_{2\beta^\prime}\sin\xi^\prime e^{-i\xi^\prime}(\cos\xi^\prime+ic_{2\beta^\prime}\sin\xi^\prime)\neq 0$ [after making use of \eq{zeesix}]. In all cases, the scalar potential and vacuum are CP-conserving.
 \label{tabalign3}}
\begin{tabular}{|c|c|c|c|c|}
\hline
$\beta^\prime$ & $\xi^\prime$ & $m_{11}^{\prime\,2}$, $m_{22}^{\prime\,2}$ & $\pht m_{12}^{\prime\,2}\pht$ & comment \Bstrut \\
\hline
$\pht s_{2\beta^\prime}c_{2\beta^\prime}\neq 0 \pht$ & $\pht\sin 2\xi^\prime\neq 0\pht$  &  $\pht m_{11}^{\prime\,2}\neq m_{22}^{\prime\,2}\pht$ &  \pht complex ($\neq 0$) \pht  & \Tstrut  \\
$\pht s_{2\beta^\prime}c_{2\beta^\prime}\neq 0 \pht$ & $\pht\sin 2\xi^\prime\neq 0\pht$  &  $\pht m_{11}^{\prime\,2}\neq m_{22}^{\prime\,2}\pht$ &  \pht real ($\neq 0$) \pht  & \pht $m_A^2=\lambda_5^\prime v^2$ \pht  \\
$s_{2\beta^\prime}c_{2\beta^\prime}\neq 0$  & $\cos \xi^\prime=0$ & $m_{11}^{\prime\,2}\neq m_{22}^{\prime\,2}$  & \pht pure imaginary ($\neq 0$) \pht  &\\
$s_{2\beta^\prime}c_{2\beta^\prime}\neq 0$  & $\cos \xi^\prime=0$ & $m_{11}^{\prime\,2}\neq m_{22}^{\prime\,2}$  & 0 & \pht $m_A^2=\lambda_5^\prime v^2$ \pht  \\
$s_{2\beta^\prime}c_{2\beta^\prime}\neq 0$  & $\sin \xi^\prime\neq 0$ & $m_{11}^{\prime\,2}= m_{22}^{\prime\,2}$  &  \pht pure imaginary ($\neq 0$) \pht  & \pht $m_A^2=0$ \pht  \\
$c_{2\beta^\prime}=0$  & $\sin 2\xi^\prime\neq 0$ & $m_{11}^{\prime\,2}=m_{22}^{\prime\,2}$  &  \pht pure imaginary ($\neq 0$) \pht & $m_A^2=0$  \\
$c_{2\beta^\prime}=0$  & $\sin 2\xi^\prime\neq 0$ & $m_{11}^{\prime\,2}=m_{22}^{\prime\,2}$  &\pht real ($\neq 0$) \pht  & \pht $m_A^2=\lambda_5^\prime v^2$ \pht  \\
$c_{2\beta^\prime}=0$  & $\sin 2\xi^\prime\neq 0$ & $m_{11}^{\prime\,2}=m_{22}^{\prime\,2}$  &\pht complex ($\neq 0$) \pht & \pht $m_A^2\neq 0$, $\lambda_5^\prime v^2$ \pht \Bstrut \\
\hline
\end{tabular}
\end{table}

In light of the results exhibited in the three tables above, we have successfully achieved natural approximate Higgs alignment in the bosonic sector of the 2HDM.   Of course,
for a truly successful model of natural approximate Higgs alignment, one must extend the symmetries of the ERPS to the Yukawa interactions.   First, consider a model with one generation of quarks and leptons.  One can quickly conclude that the Yukawa sector of a one generation model does not respect any of the symmetries of the ERPS.  That is, the Yukawa interactions (which involve dimension-four interaction terms) constitute a hard breaking of the ERPS symmetries, which necessarily spoils the naturalness of approximate Higgs alignment as one cannot maintain small symmetry breaking squared-mass parameters without fine tuning.   

In a model with three generations of quarks and leptons, one can extend the ERPS symmetries to the Yukawa sector by making use of the quark and lepton flavor degrees of freedom in defining the symmetry transformations of the fermion fields.  A comprehensive attempt 
to construct Yukawa interactions that respect a GCP2 or GCP3 symmetry was presented in Ref.~\cite{Ferreira:2010bm}.  Unfortunately, none of the resulting models were phenomenologically viable, either due to the presence of a massless fermion or (in one case) due to insufficient CP violation, with a corresponding Jarlskog invariant~\cite{Jarlskog:1985ht,Jarlskog:1985cw} that was nearly three orders of magnitude below the experimental data.

In Ref.~\cite{Draper:2020tyq}, a different strategy was employed.  To extend the GCP2 and GCP3 symmetries to the Yukawa sector, 
vector-like top (and bottom) quark partners were added to the Standard Model. These symmetries are then broken softly by vector-like quark mass parameters, thereby providing a mechanism for generating 
 the soft symmetry breaking, $m_{11}^2\neq m_{22}^2$ and $m_{12}^2\neq 0$, exhibited in  Tables~\ref{tabalign1}, \ref{tabalign2} and \ref{tabalign3}.
 A simple model that illustrates this strategy is briefly treated in the next section.
  
 \section{A softly-broken U(1)$_{\rm PQ}\otimes\Pi_2$-symmetric 2HDM with vector-like fermions}

 \label{model}
 
 The 2HDM with a GCP3-symmetric scalar potential can be realized in another scalar field basis as 
a  ${\rm U}(1)_{\rm PQ}\otimes\Pi_2$ symmetry, where [cf.~Table~\ref{symm}]
\beq
m_{11}^2=m_{22}^2\,,\quad \lambda_1=\lambda_2\,,\quad  m_{12}^2=\lambda_5=\lambda_6=\lambda_7=0\,.
\eeq
To extend this symmetry to the Yukawa sector, we introduce vector-like fermions $U$ and $\widebar{U}$~\cite{Draper:2016cag,Draper:2020tyq}.
Two-component SM fermions~\cite{Dreiner:2008tw} are denoted by lower case letters (e.g.~doublet fields $q=(u,d)$ with hypercharge $Y=1/3$ and singlet fields $\bar{u}$ with hypercharge $Y_{\bar{u}}=-4/3$), and vector-like singlet two-component fermions are denoted by upper case letters.  Note that $Y_{\bar{u}}=Y_{\widebar{U}}=-Y_U$.
Under the U(1)$_{\rm PQ}$ and $\Pi_2$ symmetries, the fields transform as shown in Table~\ref{transform} below.\footnote{The down-type quarks and leptons can also be included by introducing the corresponding vector-like fermion partners~\cite{Draper:2020tyq}, in which case the Type-I, II, X or Y Higgs-fermion Yukawa couplings~\cite{Hall:1981bc,Barger:1989fj,Aoki:2009ha} can be realized.}
\begin{table}[ht!]
\centering
\caption{Transformation of the scalar and fermion fields under the ${\rm U}(1)\otimes\Pi_2$ symmetry.\label{transform}}
\begin{tabular}{|c|cccccc|}
\hline
symmetry &  $\Phi_1$ & $\phm\Phi_2$ &
 $\phm q$ & $\phm \bar{u}$ & $\widebar{U}$ & $U$ \TBstrut  \\
\hline
$\Pi_2$  & $ \Phi_2$ & $\phm\Phi_1$ &
    $\phm q$ & $\phm \widebar{U}$ & $\bar{u}$ &  $U$\Tstrut \\[3pt]
U(1)$_{\rm PQ}$ &  $e^{-i\theta}\Phi_1$ & $e^{i\theta}\Phi_2$
 & $\phm q$ & $e^{-i\theta} \bar{u}$& $e^{-i\theta}\widebar{U}$ &  $e^{\pm i\theta}U$ \Bstrut \\
 \hline
\end{tabular}
\end{table}

The Yukawa couplings consistent with the U(1)$_{\rm PQ}\otimes\Pi_2$ symmetry and the SU(2)$\times$U(1)$_{\rm Y}$ gauge symmetry are
\beq
\mathscr{L}_{\rm Yuk} \, \supset \, y_t \left(q \Phi_2 \bar{u} + q \Phi_1 \widebar{U} \right) + \rm{h.c.}
\eeq
However, this model is not phenomenologically viable due to 
the experimental limits on vector-like fermion masses and the
existence of a massless Goldstone scalar if the U(1)$_{\rm PQ}$ symmetry is spontaneously broken.
These problems are easily avoided by introducing
SU(2)$\times$U(1)$_{\rm Y}$ preserving mass terms,
\beq \label{symmbreak}
\mathscr{L}_{\rm mass}\,\supset M_U \widebar{U}U+M_u \bar{u}U+{\rm h.c.}
\eeq
The U(1)$_{\rm PQ}$ symmetry is explicitly broken if $M_U M_u\neq 0$, whereas the $\Pi_2$ discrete symmetry is explicitly broken if $M_U\neq M_u$. 
Note that the symmetry breaking is soft, so that
corrections to the scalar potential squared-mass parameters are
protected from quadratic sensitivity to the cutoff scale $\Lambda$ of the theory.

The mass terms introduced in \eq{symmbreak} are also responsible for mixing between the top quark and its vector-like top partners.   
It is convenient to introduce the following two parameters:
\beq
M_T\equiv (M_U^2+M_u^2)^{1/2}\,,\qquad\quad \tan\gamma\equiv M_u/M_U\,.
\eeq
After electroweak symmetry breaking, the resulting fermion mass matrices can be diagonalized.  Ultimately, the top sector mixing is governed by
the parameters $\gamma$, $\tan\beta$, $y_t$ and $M_T$~\cite{Arhrib:2016rlj,Draper:2020tyq}.

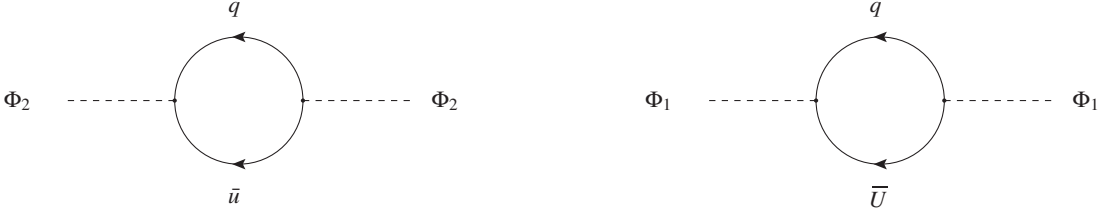
\begin{figure}[t!]
\scalebox{0.8}{
\begin{picture}(100,100)(-40,-25)
\thicklines
\DashLine(0,25)(50,25){3}
\ArrowArc(80,25)(30,0,180)
\ArrowArcn(80,25)(30,0,180)
\DashLine(110,25)(160,25){3}
\Vertex(110,25){1}
\Vertex(50,25){1}
\Text(-30,25)[l]{$\Phi_2$}
\Text(170,25)[l]{$\Phi_2$}
\Text(75,68)[l]{$q$}
\Text(75,-20)[l]{$\bar{u}$}
\DashLine(300,25)(350,25){3}
\ArrowArc(380,25)(30,0,180)
\ArrowArcn(380,25)(30,0,180)
\DashLine(410,25)(460,25){3}
\Vertex(410,25){1}
\Vertex(350,25){1}
\Text(270,25)[l]{$\Phi_1$}
\Text(470,25)[l]{$\Phi_1$}
\Text(375,68)[l]{$q$}
\Text(375,-20)[l]{$\widebar{U}$}
\end{picture}
}
\caption{One-loop contributions to $\Delta m^2\equiv m_{11}^2-m_{22}^2$.\label{loops}}
\end{figure}

One can estimate the contributions to $\Delta m^2\equiv m_{22}^2-m_{11}^2$ and $m_{12}^2$ due to the presence of the U(1)$_{\rm PQ}$$\times\Pi_2$ symmetry breaking mass terms
given in \eq{symmbreak}.  For example, corrections to $\Delta m^2=0$ arise at one-loop from the diagrams exhibited in Fig.~\ref{loops}.
An explicit computation yields~\cite{Draper:2020tyq},
\beq
\Delta m^2 \equiv m_{22}^2-m_{11}^2\sim \kappa(M_U^2-M_u^2)
-\frac{3y_t^2(M_U^2-M_u^2)}{4\pi^2}\ln(\Lambda/M_T)\,.
\eeq
The above result
includes a finite threshold correction proportional to $\kappa$.  Due to the soft nature of the U(1)$_{\rm PQ}$$\times\Pi_2$ symmetry breaking, we see that
$\Delta m^2$ depends logarithmically on the cutoff scale $\Lambda$.  Moreover, if 
$M_U=M_u$, then the $\Pi_2$ symmetry is unbroken and the relation $m_{11}^2=m_{22}^2$ is protected.
Likewise, a similar analysis of one-loop induced $\Phi_1$--$\Phi_2$ mixing yields~\cite{Draper:2020tyq},
\beq
m_{12}^2\sim\kappa_{12}M_U M_u+\frac{3y_t^2 M_U M_u}{4\pi^2}\ln(\Lambda/M_T)\,,
\eeq
which includes a finite threshold correction proportional to $\kappa_{12}$.   
Once again, $m_{12}^2$ depends logarithmically on the cutoff scale $\Lambda$.  Moreover, if $M_U M_u=0$, then the U(1)$_{\rm PQ}$ symmetry is
unbroken and the relation $m_{12}^2=0$ is protected.

We proceed to scan over the parameter space to see whether regions of approximate Higgs alignment without decoupling survive~\cite{Draper:2020tyq}.
In our numerical scans we chose $\ln(\Lambda/M_T)=3$ and $M_T=1.5$~TeV, and two benchmark points, $\gamma=0.1$ and
$\gamma=0.3$, were examined subject to the following phenomenological constraints:
\begin{itemize}
\item
Existence of a SM-like Higgs boson with mass $m_h\simeq 125$ GeV, consistent with LHC Higgs data~\cite{ATLAS:2022vkf,CMS:2022dwd}.
\item 
Heavier Higgs bosons in the parameter regime of Higgs alignment without decoupling should have so far evaded LHC detection~\cite{ATLAS:2020zms,ATLAS:2020gxx,ATLAS:2020tlo,CMS:2019ogx,CMS:2019pzc,CMS:2022goy}.
\item
Constraints on the charged Higgs mass from flavor constraints in the Type-I 2HDM~\cite{Arbey:2017gmh}.
\item
Constraints on the vector-like top quark masses  bounds and mixing parameters based on nonobservation of vector-like top quarks in LHC searches~\cite{ATLAS:2018dyh,ATLAS:2022ozf,CMS:2019afi,CMS:2022yxp}. 
\end{itemize}

Finally, note that although there is only logarithmic sensitivity to the cutoff scale $\Lambda$, one cannot take it arbitrarily large without an
excessively large fine-tuning of the parameters required to keep the corrections to $\Delta m^2$ and $m_{12}^2$ small (such that the Higgs
alignment limit is approximately realized).   For the same reason, one cannot take the vector-like top mass parameters too large.  This provided motivation
for our choice of cutoff scale of $\Lambda\simeq 30$~TeV and $M_T=1.5$~TeV as noted above.

\begin{figure}[t!]
     \centering
         \includegraphics[width=0.48\textwidth]{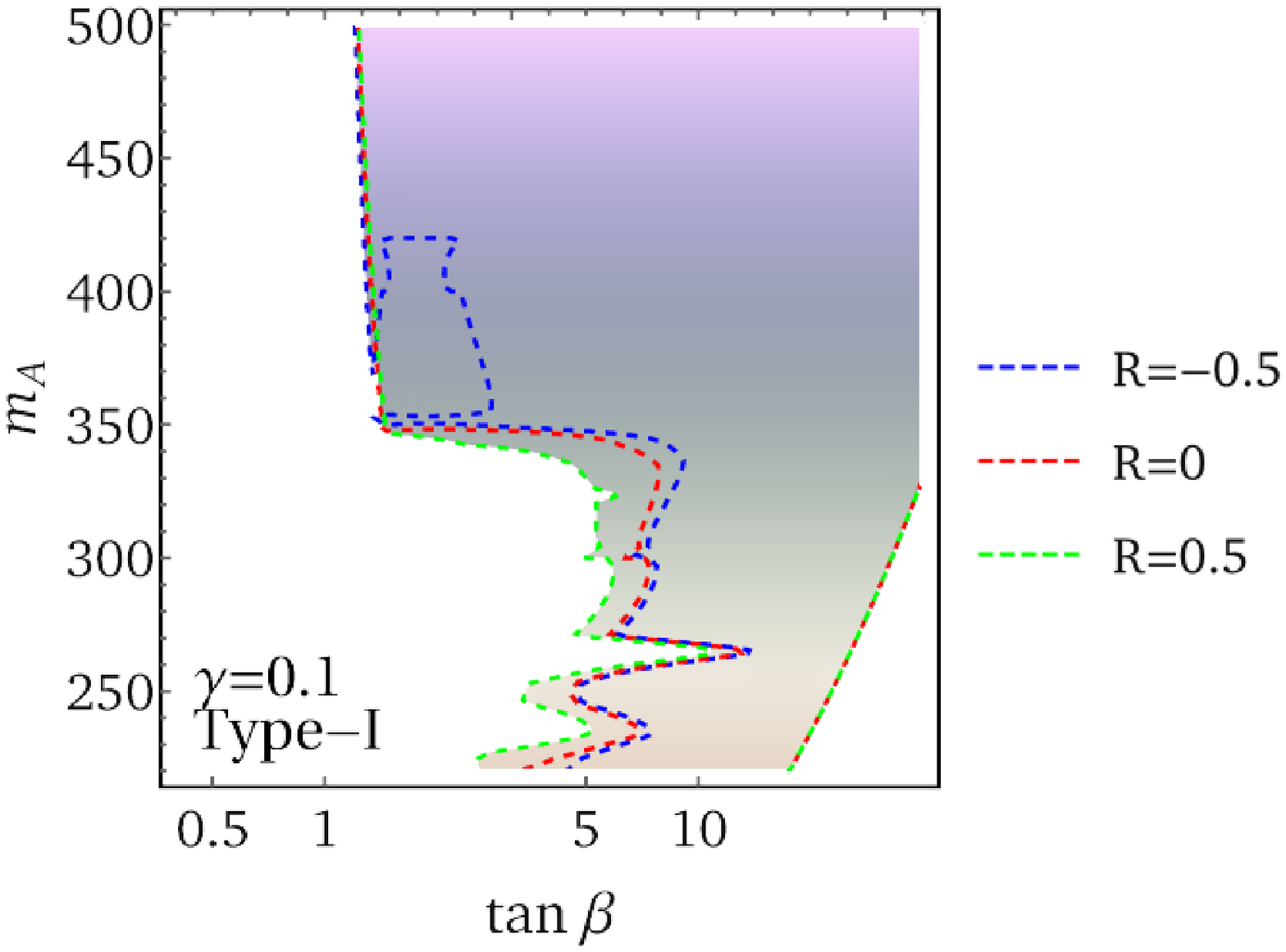}
            \includegraphics[width=0.48\textwidth]{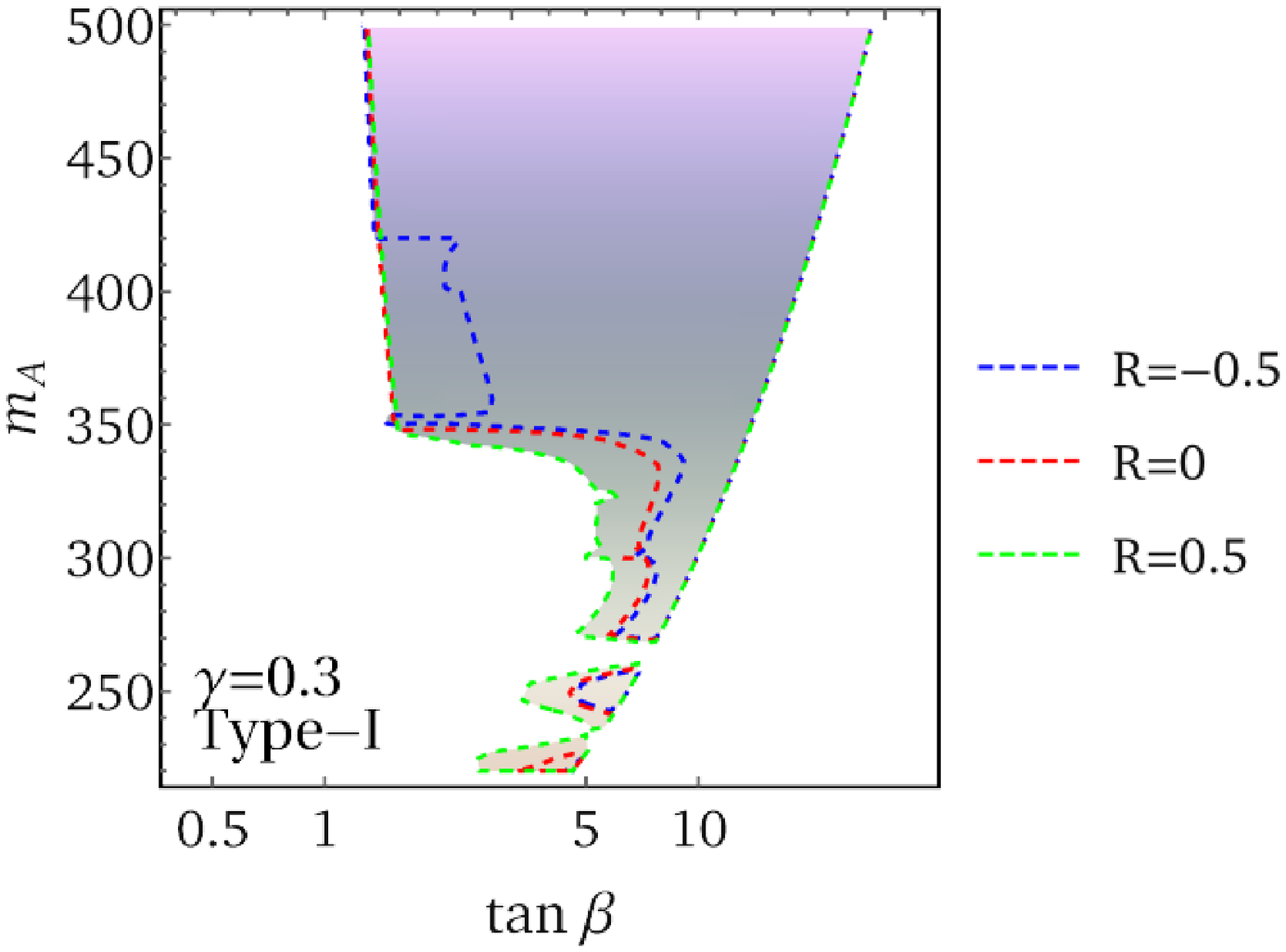} 
   \caption{ \footnotesize
    Regions allowed by experimental bounds and tuning constraints for different values of $R\equiv(\lambda_3+\lambda_4)/\lambda$, with an $m_{12}^2$ and $\Delta m^2$ tuning of at most $5\%$ [assuming that $\ln(\Lambda/M_T)=3$]. 
		 The two panels show three different $R$ curves; the white regions of the parameter space are ruled out.  The ruled out areas expand somewhat as $R$ decreases, with the borders of the allowed shaded regions indicated by the corresponding contours.  For $R=-0.5$, the area enclosed by
	the closed dashed blue contour in panel (a) is also ruled out.
Type-I Yukawa couplings are employed and, two choices for $\gamma$ are shown.  The shrinking of the allowed parameter space as $\gamma$ increases is due primarily to the behavior of the measure of fine-tuning of the parameter $m_{12}^2$.  Taken from Ref.~\cite{Draper:2020tyq}.\label{figure}}
 \end{figure}
 
In Fig.~\ref{figure}, two results from the parameter scans of Ref.~\cite{Draper:2020tyq} are exhibited. These results show that regions of approximate Higgs alignment without decoupling remain phenomenologically viable.
In particular, new scalars beyond the SM Higgs boson with masses below 500 GeV can be present with opportunities for future discovery at the LHC.  Moreover, if the framework presented here is realized in nature, then vector-like top partners must exist with masses that are not much larger than a few TeV, which again presents opportunities for discovery in future runs at the LHC.

\section{Discussion and Conclusions}

In this talk, I have presented a framework for realizing an approximate Higgs alignment without decoupling.  This framework relies on first finding a symmetry in which the Higgs alignment is exact and then breaking the symmetry softly to generate small deviations in the tree-level couplings of a SM-like Higgs boson from their Standard Model values.
In Section~\ref{model}, a low-energy effective theory (valid up to a cutoff scale of roughly 30 TeV) was exhibited in which soft-symmetry breaking terms in the Yukawa sector provide the source for the symmetry-breaking squared mass terms of the scalar potential, which are necessary for the generation of small deviations from SM-like Higgs behavior.   It would be an interesting exercise to find an ultraviolet completion of the model, which could provide insight into the origin of the ERPS symmetries and their soft breakings.

In contrast to the approach taken in this talk, there is a different strategy for achieving approximate Higgs alignment naturally, which has been advocated in
Refs.~\cite{Dev:2014yca,Dev:2017org,Pilaftsis:2016erj,Darvishi:2019ltl,Darvishi:2019dbh,Darvishi:2020teg,Darvishi:2021txa}.
In this approach, one imposes the relevant symmetries of the scalar potential at a very high energy scale~$M_X$ (e.g., the Planck scale).
However, in this approach one must accept the presence of hard symmetry breaking terms arising from the Yukawa sector.  The effect of these hard symmetry breaking terms is to modify the parameters of the scalar potential at the electroweak scale due to the contributions of the Yukawa couplings to the renormalization group running.   In this way, the size of the deviations from the exact Higgs alignment limit is controlled.  (Custodial symmetries~\cite{Sikivie:1980hm,Mannheim:1983ti},
which are not respected by the gauge covariant scalar kinetic energy terms, can be similarly treated~\cite{Pilaftsis:2011ed,Aiko:2020atr}.)
This program provides a viable alternative for generating approximate Higgs alignment without decoupling. Of course, an ultraviolet completion of the model is still necessary to explain the origin of the symmetry constraints on the scalar potential parameters that were imposed at the scale $M_X$.

If an extended Higgs sector with additional scalars exists with masses not significantly larger than the scale of electroweak symmetry breaking, then one needs to understand why the observed Higgs boson $h$ is SM-like.  
The inert doublet model (IDM) provides an example in which the tree-level properties of $h$ are exactly those of the SM, with deviations entering only via very small loop corrections.  If additional Higgs scalars are found and/or deviations of $h$ from its anticipated SM behavior (which are too large to be compatible with the IDM) are confirmed in future experiments, then a symmetry-based explanation for why the Higgs boson is SM-like could be suggesting new physics in the Yukawa sector that involves 
vector-like partners to the quarks (and leptons).

Ultimately, one of the top priorities of future collider experimentation is to answer the question of whether the Higgs sector is minimal or non-minimal, and if the latter, whether the mass scale associated with the new scalars lies significantly beyond the scale of electroweak symmetry breaking.  The answer to this question will have a profound effect on addressing the fundamental nature of the Higgs boson and the origin of the hierarchy of scales from the electroweak to the Planck scale.
 
\section*{Acknowledgments}

This presentation is based on various works in collaboration with Patrick Draper, Andreas Ekstedt, Pedro Ferreira, Joshua T.~Ruderman, and Jo\~{a}o P.~Silva.  
I am especially grateful for the many enlightening discussions that contributed to the material presented in this talk.
H.E.H. is supported in part by the U.S. Department of Energy grant number DE-SC0010107.


\begin{thebibliography}{99}

\bibitem{ATLAS:2022vkf}
The ATLAS Collaboration,
%``A detailed map of Higgs boson interactions by the ATLAS experiment ten years after the discovery,''
Nature \textbf{607}, 52 (2022)
[arXiv:2207.00092 [hep-ex]].

\bibitem{CMS:2022dwd}
The CMS Collaboration,
%``A portrait of the Higgs boson by the CMS experiment ten years after the discovery,''
Nature \textbf{607}, 60 (2022)
[arXiv:2207.00043 [hep-ex]].

\bibitem{Haber:1989xc}
H.E.~Haber and Y.~Nir,
%``Multiscalar Models With a High-Energy Scale,''
Nucl. Phys. B \textbf{335}, 363 (1990).

\bibitem{Gunion:2002zf}
J.F.~Gunion and H.E.~Haber,
%``The CP conserving two Higgs doublet model: The Approach to the decoupling limit,''
Phys. Rev. D \textbf{67}, 075019 (2003)
[arXiv:hep-ph/0207010].

\bibitem{Haber:2006ue}
H.E.~Haber and D.~O'Neil,
%``Basis-independent methods for the two-Higgs-doublet model. II. The Significance of tan$\beta$,''
Phys. Rev. D \textbf{74}, 015018 (2006)
[erratum:~Phys. Rev. D \textbf{74}, 059905 (2006)]
[arXiv:hep-ph/0602242].

\bibitem{Craig:2022uua}
N.~Craig,
``Naturalness: A Snowmass White Paper,''
[arXiv:2205.05708 [hep-ph]].

\bibitem{Ginzburg:2001wj}
I.F.~Ginzburg, M.~Krawczyk and P.~Osland,
%``Potential of photon collider in resolving SM like scenarios,''
Nucl. Instrum. Meth. A \textbf{472}, 149 (2001)
[arXiv:hep-ph/0101229].

\bibitem{Craig:2013hca}
N.~Craig, J.~Galloway and S.~Thomas,
%``Searching for Signs of the Second Higgs Doublet,''
arXiv:1305.2424 [hep-ph].

\bibitem{Asner:2013psa}
  D.M.~Asner, T.~Barklow, C.~Calancha, K.~Fujii, N.~Graf, H.E.~Haber, A.~Ishikawa, S.~Kanemura, S.~Kawada and M.~Kurata, et al.,
``ILC Higgs White Paper,''
  arXiv:1310.0763 [hep-ph].
  
\bibitem{Carena:2013ooa}
  M.~Carena, I.~Low, N.R.~Shah and C.E.M.~Wagner,
%``Impersonating the Standard Model Higgs Boson:~Alignment without Decoupling,''
  JHEP {\bf 1404} (2014) 015
  [arXiv:1310.2248 [hep-ph]].
  
\bibitem{Haber:2013mia}
  H.E.~Haber, in Proceedings of the of the Toyama International
  Workshop on Higgs as a Probe of New Physics 2013 (HPNP2013),  
%``The Higgs data and the Decoupling Limit,''
  arXiv:1401.0152 [hep-ph].

\bibitem{Carena:2015moc}
M.~Carena, H.E.~Haber, I.~Low, N.R.~Shah and C.E.M.~Wagner,
%``Alignment limit of the NMSSM Higgs sector,''
Phys. Rev. D \textbf{93}, 035013 (2016)
[arXiv:1510.09137 [hep-ph]].

\bibitem{Georgi:1978ri}
H.~Georgi and D.V.~Nanopoulos,
%``Suppression of Flavor Changing Effects From Neutral Spinless Meson Exchange in Gauge Theories,''
Phys. Lett. B \textbf{82}, 95 (1979).

\bibitem{Lavoura:1994yu}
L.~Lavoura,
%``Signatures of discrete symmetries in the scalar sector,''
Phys. Rev. D \textbf{50}, 7089 (1994)
[arXiv:hep-ph/9405307].

\bibitem{Lavoura:1994fv}
L.~Lavoura and J.P.~Silva,
%``Fundamental CP violating quantities in a SU(2) x U(1) model with many Higgs doublets,''
Phys. Rev. D \textbf{50}, 4619 (1994)
[arXiv:hep-ph/9404276].

\bibitem{Botella:1994cs}
F.J.~Botella and J.~P.~Silva,
%``Jarlskog - like invariants for theories with scalars and fermions,''
Phys. Rev. D \textbf{51}, 3870 (1995)
[arXiv:hep-ph/9411288].


\bibitem{Branco:1999fs}
G.C.~Branco, L.~Lavoura and J.P.~Silva,
\textit{CP Violation} (Oxford University Press, Oxford, UK, 1999).

\bibitem{Branco:2011iw}
G.C.~Branco, P.M.~Ferreira, L.~Lavoura, M.N.~Rebelo, M.~Sher and J.P.~Silva,
%``Theory and phenomenology of two-Higgs-doublet models,''
Phys. Rept. \textbf{516},~1 (2012)
[arXiv:1106.0034 [hep-ph]].


\bibitem{Boto:2020wyf}
R.~Boto, T.V.~Fernandes, H.E.~Haber, J.C.~Rom\~ao and J.P.~Silva,
%``Basis-independent treatment of the complex 2HDM,''
Phys. Rev. D \textbf{101}, 055023 (2020)
[arXiv:2001.01430 [hep-ph]].

\bibitem{Davidson:2005cw}
S.~Davidson and H.E.~Haber,
%``Basis-independent methods for the two-Higgs-doublet model,''
Phys. Rev. D \textbf{72}, 035004 (2005)
[erratum: Phys. Rev. D \textbf{72}, 099902 (2005)]
[arXiv:hep-ph/0504050].

\bibitem{Hall:1981bc}
L.J.~Hall and M.B.~Wise,
%``FLAVOR CHANGING HIGGS - BOSON COUPLINGS,''
Nucl. Phys. B \textbf{187}, 397 (1981).

\bibitem{Barbieri:2006dq}
R.~Barbieri, L.J.~Hall and V.S.~Rychkov,
%``Improved naturalness with a heavy Higgs: An Alternative road to LHC physics,''
Phys. Rev. D \textbf{74}, 015007 (2006)
[arXiv:hep-ph/0603188].

\bibitem{LopezHonorez:2006gr}
L.~Lopez Honorez, E.~Nezri, J.F.~Oliver and M.H.G.~Tytgat,
%``The Inert Doublet Model: An Archetype for Dark Matter,''
JCAP \textbf{02}, 028 (2007)
[arXiv:hep-ph/0612275].

\bibitem{Cirelli:2005uq}
M.~Cirelli, N.~Fornengo and A.~Strumia,
%``Minimal dark matter,''
Nucl. Phys. B \textbf{753}, 178 (2006)
[arXiv:hep-ph/0512090].

\bibitem{Ma:2006km}
E.~Ma,
%``Verifiable radiative seesaw mechanism of neutrino mass and dark matter,''
Phys. Rev. D \textbf{73}, 077301 (2006)
[arXiv:hep-ph/0601225].

\bibitem{Hambye:2009pw}
T.~Hambye, F.S.~Ling, L.~Lopez Honorez and J.~Rocher,
%``Scalar Multiplet Dark Matter,''
JHEP \textbf{07}, 090 (2009)
[erratum:~JHEP \textbf{05}, 066 (2010)]
[arXiv:0903.4010 [hep-ph]].

\bibitem{Arina:2009um}
C.~Arina, F.S.~Ling and M.H.G.~Tytgat,
%``IDM and iDM or The Inert Doublet Model and Inelastic Dark Matter,''
JCAP \textbf{10}, 018 (2009)
[arXiv:0907.0430 [hep-ph]].

\bibitem{Goudelis:2013uca}
A.~Goudelis, B.~Herrmann and O.~St\r{a}l,
%``Dark matter in the Inert Doublet Model after the discovery of a Higgs-like boson at the LHC,''
JHEP \textbf{09} (2013) 106 
[arXiv:1303.3010 [hep-ph]].

\bibitem{Bernon:2015qea}
J.~Bernon, J.F.~Gunion, H.E.~Haber, Y.~Jiang and S.~Kraml,
%``Scrutinizing the alignment limit in two-Higgs-doublet models: m$_h$=125  GeV,''
Phys. Rev. D \textbf{92}, 075004 (2015)
[arXiv:1507.00933 [hep-ph]].

\bibitem{Bernon:2015wef}
J.~Bernon, J.F.~Gunion, H.E.~Haber, Y.~Jiang and S.~Kraml,
%``Scrutinizing the alignment limit in two-Higgs-doublet models. II. m$_H$=125  GeV,''
Phys. Rev. D
\textbf{93}, 035027 (2016)
[arXiv:1511.03682 [hep-ph]].

%\cite{Hou:2017hiw}
\bibitem{Hou:2017hiw}
W.S.~Hou and M.~Kikuchi,
%``Approximate Alignment in Two Higgs Doublet Model with Extra Yukawa Couplings,''
EPL \textbf{123}, 11001 (2018)
[arXiv:1706.07694 [hep-ph]].

\bibitem{Grzadkowski:2018ohf}
B.~Grzadkowski, H.E.~Haber, O.M.~Ogreid and P.~Osland,
%``Heavy Higgs boson decays in the alignment limit of the 2HDM,''
JHEP \textbf{12}, 056 (2018)
[arXiv:1808.01472 [hep-ph]].

\bibitem{Aiko:2020ksl}
M.~Aiko, S.~Kanemura, M.~Kikuchi, K.~Mawatari, K.~Sakurai and K.~Yagyu,
%``Probing extended Higgs sectors by the synergy between direct searches at the LHC and precision tests at future lepton colliders,''
Nucl. Phys. B \textbf{966}, 115375 (2021)
[arXiv:2010.15057 [hep-ph]].

\bibitem{Kanemura:2021dez}
S.~Kanemura, M.~Takeuchi and K.~Yagyu,
%``Probing double-aligned two-Higgs-doublet models at the LHC,''
Phys. Rev. D \textbf{105}, 115001 (2022)
[arXiv:2112.13679 [hep-ph]].

\bibitem{Haber:2021zva}
H.E.~Haber and J.P.~Silva,
%``Exceptional regions of the 2HDM parameter space,''
Phys. Rev. D \textbf{103}, 115012 (2021) [erratum:~Phys. Rev. D \textbf{105}, 119902 (2022)]
[arXiv:2102.07136 [hep-ph]].

\bibitem{Dev:2014yca}
P.S.~Bhupal Dev and A.~Pilaftsis,
%``Maximally Symmetric Two Higgs Doublet Model with Natural Standard Model Alignment,''
JHEP \textbf{12}  (2014) 024
[erratum:~JHEP \textbf{11}, 147 (2015)]
[arXiv:1408.3405 [hep-ph]].

\bibitem{Dev:2017org}
P.S.~Bhupal Dev and A.~Pilaftsis,
%``Natural Alignment in the Two Higgs Doublet Model,''
J. Phys. Conf. Ser. \textbf{873}, 012008 (2017)
[arXiv:1703.05730 [hep-ph]].

\bibitem{Pilaftsis:2016erj}
A.~Pilaftsis,
%``Symmetries for standard model alignment in multi-Higgs doublet models,''
Phys. Rev. D \textbf{93}, 075012 (2016)
[arXiv:1602.02017 [hep-ph]].

\bibitem{Benakli:2018vqz}
K.~Benakli, M.D.~Goodsell and S.L.~Williamson,
%``Higgs alignment from extended supersymmetry,''
Eur. Phys. J. C \textbf{78}, 658 (2018)
[arXiv:1801.08849 [hep-ph]].

\bibitem{Benakli:2018vjk}
K.~Benakli, Y.~Chen and G.~Lafforgue-Marmet,
%``R-symmetry for Higgs alignment without decoupling,''
Eur. Phys. J. C \textbf{79}, 172 (2019)
[arXiv:1811.08435 [hep-ph]].

\bibitem{Lane:2018ycs}
K.~Lane and W.~Shepherd,
%``Natural stabilization of the Higgs boson\textquoteright{}s mass and alignment,''
Phys. Rev. D \textbf{99}, 055015 (2019)
[arXiv:1808.07927 [hep-ph]].

\bibitem{Eichten:2021qbm}
E.~Eichten and K.~Lane,
%``Higgs alignment and the top quark,''
Phys. Rev. D \textbf{103}, 115022 (2021)
[arXiv:2102.07242 [hep-ph]].


\bibitem{Peccei:1977hh}
R.D.~Peccei and H.R.~Quinn,
%``Constraints Imposed by CP Conservation in the Presence of Instantons,''
Phys. Rev. D \textbf{16}, 1791 (1977).


\bibitem{Ivanov:2007de}
I.P.~Ivanov,
%``Minkowski space structure of the Higgs potential in 2HDM. II. Minima, symmetries, and topology,''
Phys. Rev. D \textbf{77}, 015017 (2008)
[arXiv:0710.3490 [hep-ph]].

\bibitem{Ferreira:2009wh}
P.M.~Ferreira, H.E.~Haber and J.P.~Silva,
%``Generalized CP symmetries and special regions of parameter space in the two-Higgs-doublet model,''
Phys. Rev. D \textbf{79}, 116004 (2009)
[arXiv:0902.1537 [hep-ph]].

\bibitem{Ferreira:2010yh}
P.M.~Ferreira, H.E.~Haber, M.~Maniatis, O.~Nachtmann and J.P.~Silva,
%``Geometric picture of generalized-CP and Higgs-family transformations in the two-Higgs-doublet model,''
Int. J. Mod. Phys. A \textbf{26}, 769 (2011)
[arXiv:1010.0935 [hep-ph]].

\bibitem{Battye:2011jj}
R.A.~Battye, G.D.~Brawn and A.~Pilaftsis,
%``Vacuum Topology of the Two Higgs Doublet Model,''
JHEP \textbf{08} (2011) 020 
[arXiv:1106.3482 [hep-ph]].

\bibitem{tHooft:1979rat}
G.~'t Hooft,
%``Naturalness, chiral symmetry, and spontaneous chiral symmetry breaking,''
 in \textit{Recent Developments in Gauge Theories}, NATO Advanced Study Institute series: Series B, Physics; volume 59, edited by G.~'t Hooft et al. (Plenum Press, New York and London, 1980) pp.~135--157.
 
\bibitem{Ferreira:2010bm}
P.M.~Ferreira and J.P.~Silva,
%``A Two-Higgs Doublet Model With Remarkable CP Properties,''
Eur. Phys. J. C \textbf{69}, 45 (2010)
[arXiv:1001.0574 [hep-ph]].

\bibitem{Jarlskog:1985ht}
C.~Jarlskog,
%``Commutator of the Quark Mass Matrices in the Standard Electroweak Model and a Measure of Maximal $CP$~Nonconservation,''
Phys. Rev. Lett. \textbf{55}, 1039 (1985).

\bibitem{Jarlskog:1985cw}
C.~Jarlskog,
%``A Basis Independent Formulation of the Connection Between Quark Mass Matrices, CP Violation and Experiment,''
Z. Phys. C \textbf{29}, 491 (1985).

\bibitem{Draper:2020tyq}
P.~Draper, A.~Ekstedt and H.E.~Haber,
%``A natural mechanism for approximate Higgs alignment in the 2HDM,''
JHEP \textbf{05}, 235 (2021)
[arXiv:2011.13159 [hep-ph]].

\bibitem{Draper:2016cag}
P.~Draper, H.E.~Haber and J.T.~Ruderman,
%``Partially Natural Two Higgs Doublet Models,''
JHEP \textbf{06}, 124 (2016)
[arXiv:1605.03237 [hep-ph]].

\bibitem{Dreiner:2008tw}
H.K.~Dreiner, H.E.~Haber and S.P.~Martin,
%``Two-component spinor techniques and Feynman rules for quantum field theory and supersymmetry,''
Phys. Rept. \textbf{494}, 1 (2010)
[arXiv:0812.1594 [hep-ph]].

\bibitem{Barger:1989fj}
V.D.~Barger, J.L.~Hewett and R.J.N.~Phillips,
%``New Constraints on the Charged Higgs Sector in Two Higgs Doublet Models,''
Phys. Rev. D \textbf{41}, 3421 (1990).

\bibitem{Aoki:2009ha}
M.~Aoki, S.~Kanemura, K.~Tsumura and K.~Yagyu,
%``Models of Yukawa interaction in the two Higgs doublet model, and their collider phenomenology,''
Phys. Rev. D \textbf{80}, 015017 (2009)
[arXiv:0902.4665 [hep-ph]].

\bibitem{Arhrib:2016rlj}
A.~Arhrib, R.~Benbrik, S.J.D.~King, B.~Manaut, S.~Moretti and C.S.~Un,
%``Phenomenology of 2HDM with vectorlike quarks,''
Phys. Rev. D \textbf{97}, 095015 (2018)
[arXiv:1607.08517 [hep-ph]].

\bibitem{ATLAS:2020zms}
G.~Aad \text{et al.}~[ATLAS Collaboration],
%``Search for heavy Higgs bosons decaying into two tau leptons with the ATLAS detector using $pp$ collisions at $\sqrt{s}=13$ TeV,''
Phys. Rev. Lett. \textbf{125}, 051801 (2020)
[arXiv:2002.12223 [hep-ex]].

\bibitem{ATLAS:2020gxx}
G.~Aad \text{et al.}~[ATLAS Collaboration],
%``Search for a heavy Higgs boson decaying into a Z boson and another heavy Higgs boson in the $\ell \ell bb$ and $\ell \ell WW$ final states in $pp$ collisions at $\sqrt{s}=13$ $\text {TeV}$ with the ATLAS detector,''
Eur. Phys. J. C \textbf{81}, 396 (2021)
[arXiv:2011.05639 [hep-ex]].

\bibitem{ATLAS:2020tlo}
G.~Aad \text{et al.}~[ATLAS Collaboration],
%``Search for heavy resonances decaying into a pair of Z bosons in the $\ell ^+\ell ^-\ell '^+\ell '^-$ and $\ell ^+\ell ^-\nu {{\bar{\nu }}}$ final states using 139 $\mathrm {fb}^{-1}$ of proton\textendash{}proton collisions at $\sqrt{s} = 13\,$TeV with the ATLAS detector,''
Eur. Phys. J. C \textbf{81}, 332 (2021)
[arXiv:2009.14791 [hep-ex]].

\bibitem{CMS:2019ogx}
A.M.~Sirunyan \text{et al.}~[CMS Collaboration],
%``Search for new neutral Higgs bosons through the H$\to$ ZA $\to \ell^{+}\ell^{-} \mathrm{b\bar{b}}$ process in pp collisions at $\sqrt{s} =$ 13 TeV,''
JHEP \textbf{03}, 055 (2020)
[arXiv:1911.03781 [hep-ex]].

\bibitem{CMS:2019pzc}
A.M.~Sirunyan \text{et al.}~[CMS Collaboration],
%``Search for heavy Higgs bosons decaying to a top quark pair in proton-proton collisions at $\sqrt{s} =$ 13 TeV,''
JHEP \textbf{04}, 171 (2020)
[arXiv:1908.01115 [hep-ex]].

%\bibitem{CMS:2022rbd}
\bibitem{CMS:2022goy}
The CMS Collaboration,
%``Searches for additional Higgs bosons and for vector leptoquarks in $\tau\tau$ final states in proton-proton collisions at $\sqrt{s}$ = 13 TeV,''
arXiv:2208.02717 [hep-ex].

\bibitem{Arbey:2017gmh}
A.~Arbey, F.~Mahmoudi, O.~Stal and T.~Stefaniak,
%``Status of the Charged Higgs Boson in Two Higgs Doublet Models,''
Eur. Phys. J. C \textbf{78}, 182 (2018)
[arXiv:1706.07414 [hep-ph]].

\bibitem{ATLAS:2018dyh}
M.~Aaboud \textit{et al.} [ATLAS Collaboration],
%``Search for single production of vector-like quarks decaying into $Wb$ in $pp$ collisions at $\sqrt{s} = 13$ TeV with the ATLAS detector,''
JHEP \textbf{05}, 164 (2019)
[arXiv:1812.07343 [hep-ex]].

\bibitem{ATLAS:2022ozf}
G.~Aad \textit{et al.} [ATLAS Collaboration],
%``Search for single production of a vector-like $T$ quark decaying into a Higgs boson and top quark with fully hadronic final states using the ATLAS detector,''
Phys. Rev. D \textbf{105}, 092012 (2022)
[arXiv:2201.07045 [hep-ex]].

\bibitem{CMS:2019afi}
A.M.~Sirunyan \textit{et al.} [CMS Collaboration],
%``Search for electroweak production of a vector-like T quark using fully hadronic final states,''
JHEP \textbf{01}, 036 (2020)
[arXiv:1909.04721 [hep-ex]].

\bibitem{CMS:2022yxp}
A.~Tumasyan \textit{et al.} [CMS Collaboration],
%``Search for single production of a vector-like T quark decaying to a top quark and a Z boson in the final state with jets and missing transverse momentum at $ \sqrt{s} $ = 13 TeV,''
JHEP \textbf{05}, 093 (2022)
[arXiv:2201.02227 [hep-ex]].

\bibitem{Darvishi:2019ltl}
N.~Darvishi and A.~Pilaftsis,
%``Quartic Coupling Unification in the Maximally Symmetric 2HDM,''
Phys. Rev. D \textbf{99},  115014 (2019)
[arXiv:1904.06723 [hep-ph]].

\bibitem{Darvishi:2019dbh}
N.~Darvishi and A.~Pilaftsis,
%``Classifying Accidental Symmetries in Multi-Higgs Doublet Models,''
Phys. Rev. D \textbf{101}, 095008 (2020)
[arXiv:1912.00887 [hep-ph]].

\bibitem{Darvishi:2020teg}
N.~Darvishi and A.~Pilaftsis,
%``Natural Alignment in Multi-Higgs Doublet Models,''
PoS \textbf{CORFU2019}, 064 (2020)
[arXiv:2004.04505 [hep-ph]].

\bibitem{Darvishi:2021txa}
N.~Darvishi, M.R.~Masouminia and A.~Pilaftsis,
%``Maximally symmetric three-Higgs-doublet model,''
Phys. Rev. D \textbf{104}, 115017 (2021)
[arXiv:2106.03159 [hep-ph]].

 \bibitem{Sikivie:1980hm}
P.~Sikivie, L.~Susskind, M.B.~Voloshin and V.I.~Zakharov,
%``Isospin Breaking in Technicolor Models,''
Nucl. Phys. B \textbf{173}, 189 (1980).

 \bibitem{Mannheim:1983ti}
P.D.~Mannheim,
%``Effective Low-Energy Custodial Symmetry and Weinberg Mixing,''
Phys. Lett. B \textbf{125}, 282 (1983).

\bibitem{Pilaftsis:2011ed}
A.~Pilaftsis,
%``On the Classification of Accidental Symmetries of the Two Higgs Doublet Model Potential,''
Phys. Lett. B \textbf{706}, 465 (2012)
[arXiv:1109.3787 [hep-ph]].

\bibitem{Aiko:2020atr}
M.~Aiko and S.~Kanemura,
%``New scenario for aligned Higgs couplings originated from the twisted custodial symmetry at high energies,''
JHEP \textbf{02}, 046 (2021)
[arXiv:2009.04330 [hep-ph]].

\end{thebibliography}
\end{document}